\documentclass[acus]{epac98}
\usepackage{epsfig}

\title{CAVITY LOSS FACTORS FOR NON-ULTRARELATIVISTIC BEAMS}
\author{Sergey~S.~Kurennoy \\
        LANSCE-1, Los Alamos National Laboratory,
        Los Alamos, NM 87545, USA }

\begin{document}
\maketitle

\begin{abstract} 
Cavity loss factors can be easily computed for ultrarelativistic
beams using time-domain codes like MAFIA or ABCI. However, for
non-ultrarelativistic beams the problem is more complicated because
of difficulties with its numerical formulation in the time domain. 
We calculate the loss factors of a non-ultrarelativistic bunch and 
compare results with the relativistic case. 
\end{abstract} 

\section{Introduction}

It is common to believe that loss factors of a bunch moving along
an accelerator structure at velocity $v=\beta c$ with $\beta < 1$ 
are lower than those for the same bunch in the ultrarelativistic 
case, $\beta \to 1$. The loss factors are then computed numerically 
for the ultrarelativistic bunch, which is a relatively 
straightforward task, and considered as upper estimates for the 
case in question, $\beta < 1$.

We study $\beta$-dependence of loss factors in an attempt 
to develop a method to obtain answers for $\beta < 1$ case
from the results for $\beta=1$. It is demonstrated that the above 
assumption on the upper estimate might be incorrect in some cases,
depending on the bunch length and properties of the 
structure (cavity + pipe) under consideration.

\section{Beam Coupling Impedance and Loss Factors of a Cavity}

In the  frequency  domain and  in  the  "closed-cavity" approximation 
(which means very narrow beam pipes) the beam coupling impedance 
calculation can be reduced  to an internal eigenvalue boundary problem. 
Let $\vec E_s$, $\vec H_s$ be a complete set of eigenfunctions (EFs) 
for the boundary problem in a closed cavity with perfect walls.  
The longitudinal impedance is then given by (e.g., \cite{SKrev})
\begin{equation}
Z(\beta, \omega ) = -i\omega \sum_s {{1}\over{\omega ^2_s-\omega ^2}}
 {{|I_s(\beta, \omega )|^2} \over {2 W_s} }  \, ,        \label{impser}
\end{equation}
where  $I_s(\beta, \omega ) = \int_L{dz \exp (-i\omega z/\beta c)
 E_{sz}(0,z)} $ is the overlap integral, and $W_s$ is the energy stored
in the $s$-th mode. Here $E_{sz}(0,z)$ is the longitudinal component of 
the $s$-th mode electric field taken on the chamber axis. 

There is a resonant enhancement of the $s$-th term in the series 
(\ref{impser}) for $Z(\beta,\omega )$ as $\omega \to 
 \omega _s$. Let us introduce a finite, but small absorption 
into the cavity  walls by adding an imaginary part to the eigenvalue:
$\omega _s\to \omega '_s-i\omega ''_s = \omega '_s(1-i/2Q_s)$. 
Here the Q-value of the $s$-th mode is $Q_s=\omega'_s W_s/P_s \gg 1$, 
where $P_s$ is the averaged power dissipated in the cavity walls (
plus, in a real structure, due to radiation into beam pipes). 
For $\omega \simeq \omega '_s$  the $s$-th term  in
Eq.\ (\ref{impser}) dominates:   
 \begin{equation}
Z(\omega \simeq \omega '_s) \simeq R_s(\beta) = {Q_s\over 
 2 \omega '_s W_s} |I_s(\beta, \omega '_s)|^2 \, .    \label{impres}
\end{equation}
The quantity $R_s(\beta)$ is the shunt impedance of the $s$-th cavity 
mode, and, unlike the $Q$-factor, it depends on $\beta$.

The beam loss factor is 
\begin{equation}
k = \frac{1}{\pi } \int_0^{\infty}d\omega \ Re\,Z(\beta,\omega )
 |\lambda (\omega )|^2 \ ,                           \label{kloss}
\end{equation}
where  $\lambda (\omega )=\int ds\exp{[i\omega s/(\beta c)]}
 \lambda(s)$  is a harmonic of bunch spectrum. For a Gaussian bunch
with rms length $2l$, the line density is  
 $\lambda (s)=\exp{(-s^2/2l^2)}/(\sqrt{2\pi }l)$ and 
 $\lambda (\omega )= \exp \{-[\omega l/(\beta c)]^2/2) \} $. 
Assuming all $Q_s >> 1$ and integrating formally Eq.\ (\ref{impser}) 
for the $Re\,Z(\beta,\omega)$, one can express the loss factor 
as a series 
\begin{equation}
k(\beta,l) = \sum_s k_s(\beta,l) = \sum_s 
\exp{\left [-\left (\frac{\omega'_s l}{\beta c} \right )^2\right ]}
\frac {\omega'_s R_s(\beta)}{2 Q_s}  \ ,                  \label{k_s}
\end{equation}
where the loss factors of individual modes $k_s$ in the last equation 
are written for the Gaussian bunch.

In principle, Eq.\ (\ref{k_s}) give us the dependence
of the loss factor on $\beta$. However, 
the answer was obtained in the ``closed-cavity'' approximation.
Moreover, it is practical only when the number of strong resonances 
is reasonably small, since their the $\beta$-dependence varies from 
one resonance to another:
\begin{equation}
\frac{k_s(\beta,l)}{k_s(1,l)} = 
\exp{\left [-\left (\frac{\omega'_s l}{c}\right )^2 
\frac{1}{\beta^2 \gamma^2} \right ]}
\frac {R_s(\beta)}{R_s(1)}  \ ,                     \label{ratk_s}
\end{equation}
where $\gamma=1/\sqrt{1-\beta^2}$.
It is obvious from Eq.\ (\ref{ratk_s}) that for long bunches loss
factors will decrease rapidly with $\beta$ decrease, as 
$\exp \left(-\beta^{-2} \right)$. Indeed, the 
lowest resonance frequencies are $\omega'_s \approx c/d$, where 
$d$ is a typical transverse size of the cavity. The exponent 
argument $-(l/d)^2$ will have a large negative value for $l \ge d$, 
and the exponential decrease for small $\beta$ will dominate the 
impedance ratio. The impedance ratio dependence on $\beta$ is more 
complicated, and we consider below a few typical examples.

\section{Examples}
 
\subsection{Cylindrical Pill-Box}
 
For a cylindrical cavity in the limit of a vanishing radius of beam 
pipes, $b\to 0$, one can obtained explicit expressions of the mode 
frequencies and impedances, e.g., \cite{SKrev}. 
Let the cavity length be $L$ and its 
radius be $d$. The mode index $s=(m,n,p)$ means that there are $m$ 
radial variations and $p$ longitudinal ones of the mode $E$-field. 
The resonance frequency is 
$\omega _{mnp}= \sqrt{\mu ^2_{mn}+ (\pi pd/L)^2}c/d$,
where $\mu_{mn}$ is the $n$-th zero of the first-kind Bessel function 
$J_m(x)$. The longitudinal shunt impedance is
 \begin{eqnarray}
R_{0np}  =  {Z_0\over  2\pi \beta ^2}  {L^3\over  d^2  \delta } {\mu
^2_{0n}\over J^2_1(\mu _{0n})} 
{c\over \omega _{0np}d} {1\over 1+\delta _{p0}+2d/L}
\times \nonumber\\
\times \left[ \left( {\omega _{0np}L\over  2\beta c} \right)^2
- \left( {\pi p\over 2}\right)^2 \right]^{-2}
\left\{ {\sin^2 \atop \cos^2} \right\}
\left( {\omega _{0np}L\over  2\beta c} \right) .   \label{rpbl}
\end{eqnarray}
The upper line in \{$\ldots$\} corresponds to even $p$ and the lower
one to odd $p$, and $\delta $ is the skin-depth.

The ratio of loss factors Eq.\ (\ref{ratk_s}) for the lowest  
$E$-mode, $E_{010}$, is then
\begin{equation}
\frac{k_{010}(\beta,l)}{k_{010}(1,l)} = 
\exp{\left [-\left (\frac{\mu_{01} l}{d\beta\gamma} \right )^2 
 \right ]} \left ( \beta \frac { \sin{\frac{\mu_{01}L}{2\beta d}} }
 {\sin{\frac{\mu_{01}L}{2 d}}} \right )^2  .  
                   \label{ratk_c010}
\end{equation}
Obviously, it is almost independent of $\beta$ when the bunch is short, 
$l \ll d$, and the cavity is short compared to its radius, $L \ll d$. 
For longer cavities, however, the ratio oscillates and might exceed 1.
This strong resonance behavior is clearly seen in Fig.~1 for large 
$L/d$, while for small $L/d$ the $k$-ratio slowly decreases with
$\beta$ decrease. For some particular parameter values, $k_{010}(\beta)$ 
can be many times larger than $k_{010}(1)$. A picture for a longer bunch 
is similar except the resonances at small $\beta$s are damped heavily.

\begin{figure}[htb]
\centerline{\epsfig{figure=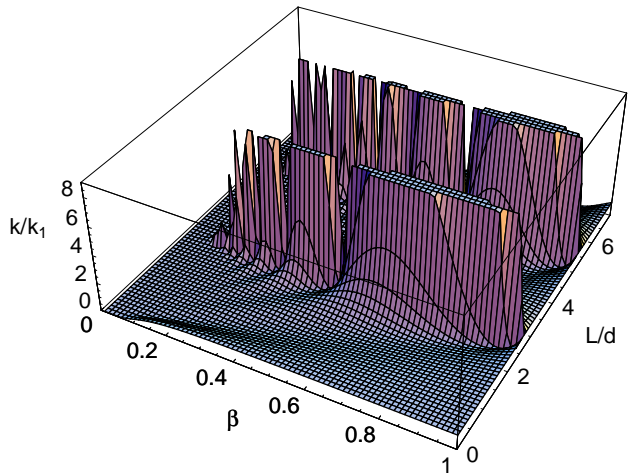,width=82.5mm}}
\caption{Ratio of loss factors (\ref{ratk_c010}) 
for a short bunch, $l/d=0.05$, versus 
$\beta$ and resonator length $L/d$.}
\end{figure}

\subsection{APT 1-cell Cavity}

As a more realistic example, we consider an APT superconducting (SC)
1-cell cavity with a power coupler \cite{FLK97}. Of course, such 
a cavity with wide beam pipes to damp higher order modes can not 
be described completely by the formalism of Sect.~2, except for the 
modes below the pipe cutoff. Direct time-domain computations with the
codes MAFIA \cite{MAFIA} and ABCI \cite{ABCI} show the existence 
of only 2 longitudinal modes below the cutoff for the $\beta=0.64$ 
cavity, and only 1 for $\beta=0.82$, in both cases 
including the fundamental mode at $f_0=700$~MHz. 
The loss factor contributions from these lowest resonance modes for 
a Gaussian bunch with the length $l=3.5$~mm for $\beta=0.64$, 
and $l=4.5$~mm for $\beta=0.82$, are about 1/3 of the total loss 
factor.

We use MAFIA results for the field of the lowest mode
to calculate the overlap integral and study the loss factor 
dependence on $\beta$. 
The on-axis longitudinal field of the fundamental mode is fitted 
very well by a simple formula $E_z(z)=E_z(0)\exp{[-(z/a)^2]}$,
where $a=0.079$~m for $\beta=0.64$ and $a=0.10$~m for $\beta=0.82$,
see \cite{SK-LACP} for detail.
The ratio of the shunt impedances in Eq.\ (\ref{ratk_s}) is then 
easy to get analytically
\begin{equation}
\frac {R_s(\beta)}{R_s(1)} =
\exp{\left [- \frac{1}{2} \left (\frac{\omega a}{c}\right )^2 
\frac{1}{\beta^2 \gamma^2} \right ]} \ ,             \label{ratR-1c}
\end{equation}
where $\omega=2\pi f_0$. The resulting dependence shows a smooth 
decrease at lower $\beta$s. 
The loss factor for the lowest mode for $\beta=0.64$ is 
0.614 times that with $\beta \to 1$, and for $\beta=0.82$ is
0.768 times the corresponding $\beta = 1$ result.

\subsection{APT Cavity, 5 cells}

For 5-cell APT SC cavities the lowest resonances are split into 5 
modes which differ by phase advance per cell $\Delta \Phi$, 
and their frequencies are a few percent apart \cite{FLK97}. 
We use MAFIA results \cite{FLK} for these modes to calculate
their loss factors according to Eq.~(\ref{k_s}). 
The on-axis fields of two modes, with $\Delta \Phi = 0$ 
(0-mode) and $\Delta \Phi = \pi$ ($\pi$-mode), which is 
the cavity accelerating mode, are shown in Fig.\ 2.

\begin{figure}[htb]
\centerline{\epsfig{figure=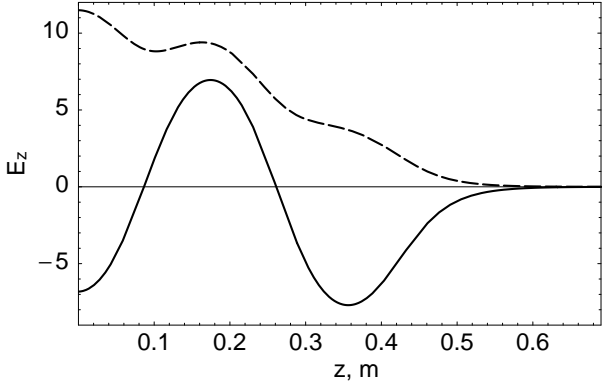,width=82.5mm}}
\caption{Longitudinal on-axis electric field 
(arbitrary units) for 0-mode (dashed) and fundamental 
($\pi$-) mode in a half of the 5-cell APT $\beta=0.82$ 
cavity.}
\end{figure}

Time-domain simulations with the code ABCI \cite{ABCI} give us 
the loss factor of a bunch at $\beta=1$.
The loss factor spectrum for the $\beta=0.64$ cavity, integrated 
up to a given frequency, has two sharp steps: one near 700~MHz 
with the height 0.5~V/pC and the other near 1400~MHz with the 
height about 0.1~V/pC. They correspond to the two bands of the 
trapped monopole modes in the cavity, cf.\ Table~1.
  
We calculate numerically overlapping integrals in Eq.~(\ref{k_s}) 
for a given $\beta$. The results for the loss factors of the 
lowest monopole modes are presented in Table~1. 
The totals for the TM$_{010}$ and TM$_{020}$ bands for $\beta=1$ 
in Table~1 agree very well with the time-domain results. In fact, 
we are mostly concerned about only these two 
resonance bands, since the higher modes are above the cutoff, and 
they propagate out of the cavity into the beam pipes depositing most 
of their energy there. Our results for the design values of $\beta$ 
are in agreement with those obtained in \cite{FLK97}. 
Remarkably, the total loss factors for a given resonance band in
Table~1 are lower for the design $\beta$ than at $\beta=1$. The
only exception is the TM$_{020}$ band for the $\beta=0.82$ cavity, 
but it includes some propagating modes, and its contribution is 
very small.

The $\beta$-dependence of the loss factor for two TM$_{010}$ 
modes mentioned above (0- and $\pi$-mode) is shown in Fig.\ 3.
Obviously, the shunt impedance (and the loss factor) dependence 
on $\beta$ is strongly influenced by the mode field pattern. 

\begin{table}[htb]
\caption{Loss Factors (in V/pC) in APT 5-cell Cavities}
\begin{tabular}{|c|c|c|c|c|} \hline
 $\Delta \Phi$ & $f$, MHz & $k(\beta)$ & $k(1)$ & $k(\beta)/k(1)$ \\  
\hline
\multicolumn{5}{|c|}{$\beta=0.64$,  TM$_{010}$-band} \\
\hline
        0 & 681.6 & $7.2 \, 10^{-6}$ & $3.7 \, 10^{-4}$ & 0.020 \\  
 $2\pi/5$ & 686.5 & $4.8 \, 10^{-5}$ &  $2.9 \, 10^{-2}$ & 0.0016 \\
 $3\pi/5$ & 692.6 & $1.1 \, 10^{-4}$ & 0.218 &  0.0005 \\  
 $4\pi/5$ & 697.6 & $1.2 \, 10^{-3}$  & 0.250 &  0.0049 \\  
 $\pi$ & 699.5 & 0.184 & $9.2 \, 10^{-3}$ & 19.92 \\  
\hline
 Total &  & 0.185 & 0.507 & 0.365 \\
\hline
\multicolumn{5}{|c|}{$\beta=0.64$,  TM$_{020}$-band} \\
\hline
    0 & 1396.8 & $6.5 \, 10^{-4}$ & $5.4 \, 10^{-4}$& 1.187 \\  
 $2\pi/5$ & 1410.7 & $1.2 \, 10^{-6}$ & $9.0 \, 10^{-4}$ & 0.0014 \\  
 $3\pi/5$ & 1432.7 & $1.8 \, 10^{-5}$  & 0.0173 &  0.0011 \\
 $4\pi/5$ & 1458.8 & $8.0 \, 10^{-7}$ & 0.0578 &$1.4 \, 10^{-5}$ \\  
 $\pi$ & 1481.0 & $3.5 \, 10^{-7}$ & 0.0095 & $3.7 \, 10^{-5}$ \\
\hline
 Total  & & $6.7\, 10^{-4}$ & 0.086 & $7.8\, 10^{-3}$ \\
\hline
\multicolumn{5}{|c|}{$\beta=0.82$, TM$_{010}$-band} \\
\hline
   0 & 674.2 & $0.3\, 10^{-6}$ & $6.9\, 10^{-4}$ & $4.5\, 10^{-4}$ \\  
 $2\pi/5$ & 681.2 & $7.3 \, 10^{-5}$ & $1.6 \, 10^{-5}$ & 4.64 \\  
 $3\pi/5$ & 689.9 & $1.8 \, 10^{-6}$ & 0.034 & $5.1\, 10^{-5}$ \\  
 $4\pi/5$ & 697.2 & $1.3 \, 10^{-3}$ & 0.220 & $5.9\, 10^{-3}$ \\  
 $\pi$ & 699.9 & 0.285 & $0.240 $ & 1.188 \\  
\hline
 Total & & 0.286 & 0.494 & 0.579 \\
\hline
\multicolumn{5}{|c|}{$\beta=0.82$, TM$_{020}$-band} \\ 
\hline
   0 & 1357.7 & $4.2 \, 10^{-5}$ & $0.8 \, 10^{-6}$ & 52.4 \\  
 $2\pi/5$ & 1367.7 & $1.4 \, 10^{-4}$ & $8.0 \, 10^{-5}$ & 1.71 \\  
 $3\pi/5$ & 1384.5 & $1.6 \, 10^{-6}$  & $1.4 \, 10^{-4}$ & 0.011 \\
 $4\pi/5 {}^{*}$ & 1409.6 & $8.0 \, 10^{-7}$  & 
   $1.3 \, 10^{-3}$ & $5.6 \, 10^{-3}$ \\  
 $\pi {}^{**}$ & 1436.9 & $1.6 \, 10^{-2}$ & 
   $2.2 \, 10^{-3}$ & 7.5 \\
\hline
 Total & & $1.6 \, 10^{-2}$ & $ 3.7 \, 10^{-3}$ & 4.32 \\
\hline
\end{tabular}
${}^{*}$Mode near the cutoff.\\
${}^{**}$Propagating mode, above the cutoff. 
\end{table}

\begin{figure}[htb]
\centerline{\epsfig{figure=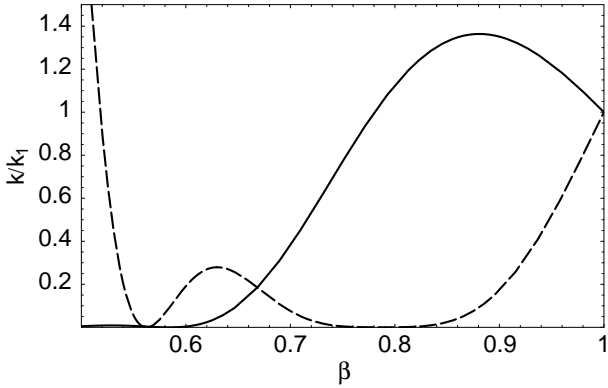,width=82.5mm}}
\caption{Loss factor ratio vs $\beta$ for 0-mode (dashed) and 
fundamental ($\pi$-) mode in the 5-cell APT $\beta=0.82$ 
cavity.}
\end{figure}

\section{Summary}

The examples above compare loss factors for $\beta < 1$ with $\beta 
 \to 1$ results. More details can be found in \cite{SK-LACP}. 
Essentially, the frequency-domain approach has been applied instead 
of the time-domain one. It can be done only when we know the fields of 
all modes contributing significantly into the loss factor. 
Nevertheless, for many practical applications, including 
SC cavities, the lowest mode contribution is a 
major concern, because propagating modes travel out of the cavity 
and deposit their energy away from the structure cold parts.

One interesting observation is that the loss factor of an individual
mode at some $\beta < 1$ can be many times larger than for 
$\beta = 1$. Obviously, one should exercise caution in using 
$\beta = 1$ results as upper estimates for a $\beta < 1$ case. 

The author would like to thank Frank Krawczyk for fruitful discussions
and for providing MAFIA results for the 5-cell cavities. 
Useful discussions with Robert Gluckstern and Thomas Wangler are 
gratefully acknowledged.

\end{document}